\documentclass[aps, prb, amsmath, reprint]{revtex4-1}
\usepackage{graphicx}
\usepackage[breaklinks=true,colorlinks=true,linkcolor=blue,citecolor=blue,urlcolor=blue,plainpages=false,pdfpagelabels=false]{hyperref}

\begin{document}
\title{Experimental demonstration of pitfalls and remedies for precise force deconvolution in frequency-modulation atomic force microscopy}
\author{Ferdinand Huber}
\author{Franz J. Giessibl}
\email{franz.giessibl@ur.de}
\affiliation{Institute of Experimental and Applied Physics, University of Regensburg, 93040 Regensburg, Germany}
\date{\today}

\begin{abstract}
Frequency-modulation atomic force microscopy provides an outstanding precision of the measurement of chemical bonding forces. However, as the cantilever oscillates with an amplitude $A$ that is usually on the order of atomic dimensions or even larger, blurring occurs. To extract a force versus distance curve from an experimental frequency versus distance spectrum, a deconvolution algorithm to recover the force from the experimental frequency shift is required. It has been recently shown that this deconvolution can be an ill-posed inversion problem causing false force-distance curves. Whether an inversion problem is well- or ill-posed is determined by two factors: the shape of the force-distance curve and the oscillation amplitude used for the measurement. A proper choice of the oscillation amplitude as proposed by the inflection point test of Sader \textit{et al.} [Nat. Nanotechnol. \textbf{13}, 1088 (2018)] should avoid ill-posedness. Here, we experimentally validate their inflection point test by means of two experimental data sets: force-distance spectra over a single carbon monoxide molecule as well as a Fe trimer on Cu(111) measured with a set of deliberately chosen amplitudes. Furthermore, we comment on typical pitfalls which are caused by the discrete nature of experimental data and provide MATLAB code which can be used by everyone to perform this test with their own data.
\end{abstract}
\maketitle

\section{Introduction}
Chemical bonding forces and -energies with their characteristic distance dependence can be measured precisely by atomic force microscopy (AFM).\cite{Binnig1986} Frequency-modulation AFM (FM-AFM),\cite{Albrecht1991} the most precise version of AFM, translates an averaged bonding force gradient into a frequency shift $\Delta f$ of a cantilever that oscillates at amplitude $A$. Precise measurements of bonding forces have been obtained for silicon in 2001, \cite{Lantz2001} and in 2007, the chemical identity of surface atoms was achieved by force spectroscopy.\cite{Sugimoto2007} These experiments were obtained with silicon cantilevers with a stiffness on the order of 10\,N/m and with relatively large oscillation amplitudes on the order of about 10\,nm. Although a study from 1999\cite{Giessibl1999} suggested that optimized signal-to-noise ratio (SNR) is obtained for oscillation amplitudes that are on the order of the decay length of the chemical interactions (around 50\,pm), large amplitudes are required for stability when using soft silicon cantilevers\cite{Giessibl1997} as in the experiments listed above.\cite{Lantz2001,Sugimoto2007} Today, many experimenters use self-detecting quartz cantilevers with a stiffness on the order of 1\,kN/m (qPlus sensors) that allow the use of small oscillation amplitudes on the order of the SNR optimizing value of about 50\,pm. Over the last decade, hundreds of notable results in modern areas of condensed matter physics have been obtained were small amplitudes were used, ranging from the measurements of forces acting during atomic manipulation,\cite{Ternes2008} the first imaging of organic molecules with atomic resolution,\cite{Gross2009,Pavlicek2017NatChemRev} topological insulators,\cite{Pielmeier2015} resolution of spin, \cite{Pielmeier2013PRL} the introduction of new inert probe tips,\cite{Moenig2018} superlubricity,\cite{Kawai2016Science} carbon nanoribbons,\cite{Ruffieux2016Nature,Kawai2018SciAdv} polarity compensation mechanisms in insulating perovskites,\cite{Setvin2018}, atomic silicon logic,\cite{Rashidi2018} the observation of transitions from physisorption to chemisorption,\cite{Huber2019} the atomically precise measurement of chemical reactivity of iron clusters \cite{Berwanger2020} and atomically resolved studies of petroleum. \cite{Fatayer2018GeoPhysResLett,Chen2020}
  Today, the benefits of small amplitude operation have even been made possible for silicon cantilevers with a stiffness also on the order of 1\,kN/m and a low-noise optical detector. \cite{Arima2016}

It is important to note that atomic force microscopy goes beyond imaging, and the experimental force versus distance data can be compared to theory, e.g. to  density functional theory.\cite{Chelikowsky2019} The precise extraction of the bonding forces from experimental frequency shifts is a sizable challenge, in particular for small oscillation amplitudes and complex bonding situations that involve multiple inflection points\cite{Sader2018} as observed in recently in transitions from physisorption to chemisorption.\cite{Huber2019}

The frequency shift $\Delta f(z)$ is analytically given by a convolution of the force $F(z)$ with a weight function in an interval set by the sensor oscillation amplitude $A$:\cite{Giessibl2001}
\begin{equation}
\label{eq:df}
\Delta f(z) = \frac{f_0}{\pi k A^2} \int_{-A}^A F(z + A - q) \frac{q}{\sqrt{A^2-q^2}} \textrm{d}q,
\end{equation}
where $z$ is the tip-sample distance of closest approach, i.e. the lower turnaround point of the tip oscillation, $f_0$ is the unpertubed resonance frequency and $k$ is the stiffness of the force sensor. In order to obtain $F(z)$ from a $\Delta f(z)$ curve, Eq. (\ref{eq:df}) must be deconvoluted or -- seen from a mathematical perspective -- inverted. Various solutions exist for this inversion: analytical methods in the limit of very small\cite{Albrecht1991} or very big oscillation amplitudes,\cite{Durig1999} iterative methods\cite{Gotsmann1999,Durig2000} and more complex techniques which requires knowledge of the amplitudes and phases of higher harmonics\cite{Durig2000a} or the frequency shift as a function of amplitude.\cite{Holscher1999} Both the Sader-Jarvis\cite{Sader2004} and matrix method of Giessibl\cite{Giessibl2001} can be used for force deconvolution with any oscillation amplitude and both methods are established and widely used.

However, a recent study by Sader \textit{et al.}\cite{Sader2018} reported that the inversion of Eq. (\ref{eq:df}) can be an ill-posed problem, i.\,e. that the recovered force may be extremely sensitive to arbitrarily small errors in the frequency shift, amplitude or $z$ values. Since the tip oscillates with a finite amplitude in FM-AFM ($A > 0$ is required to track a frequency) the force-distance behavior in the interval $\left[ z, z+2A\right]$ is blurred in the measured frequency shift signal at a single $z$ [see Eq. (\ref{eq:df})]. If the curvature of $F(z)$ changes too rapidly in that interval, information is lost in $\Delta f$ and the inversion problem becomes ill-posed. This happens at an inflection point where the curvature changes its sign. 
In that case, the recovered force curve deviates from the actual one for $z$ values smaller than the position of the inflection point. This motivated the development of a test for the validity of the force deconvolution, the \textit{so-called} inflection point test.\cite{Sader2018}

Here, we explore the inflection point test in more detail and demonstrate by means of two experimental data sets how to acquire valid force-distance curves using the inflection point test. MATLAB code that was developed in this work to test discrete data is available in the supplementary material and can be used by everyone to easily perform this test with their own data.

\section{The inflection point test}
\subsection{Theory}

\begin{figure}
	\includegraphics{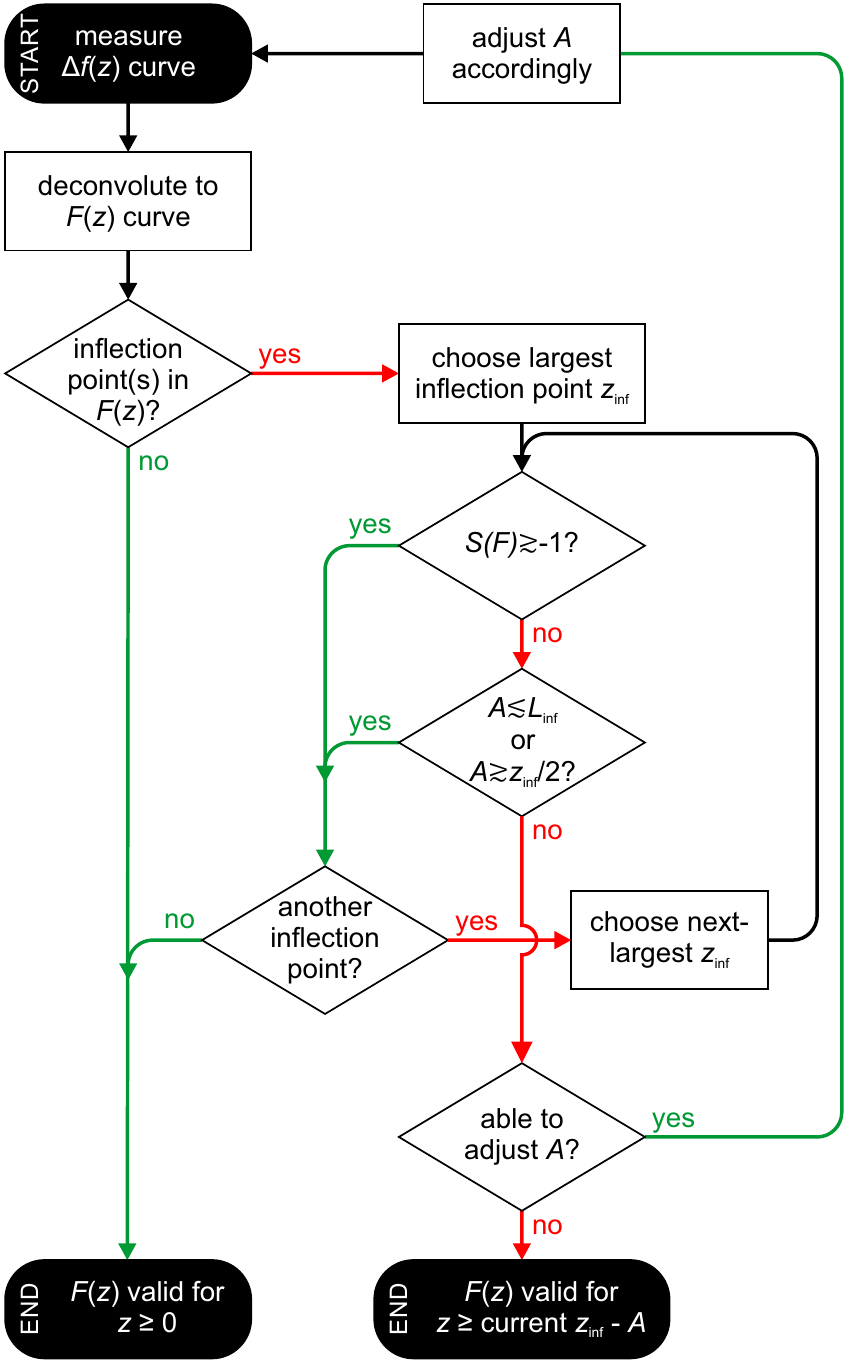}
	\caption{\label{fig:flowchart} Flow chart visualizing the inflection point test. Black round boxes indicate beginning/end, rectangular boxes operations and diamond boxes decisions, respectively. See text for detailed description and definition of all variables.} 
\end{figure}

Figure \ref{fig:flowchart} visualizes all steps of the inflection point test.\cite{Sader2018}
Starting from a $\Delta f(z)$ curve, this data is deconvoluted (with any method) to obtain the force-distance curve. The test requires that the closest tip-sample distance in this curve is defined as $z = 0$. If the $F(z)$ curve contains no inflection points the force deconvolution is valid. In case there are one or more inflection points, the test must be applied to the inflection point $z_\textrm{inf}$ with the largest $z$ value. To validate the force deconvolution for $z \gtrsim z_\textrm{inf}$ the inequality of the \textit{so-called} $S$-factor,
\begin{equation}
	\label{eq:S}
	S(F) \equiv \frac{z_\textrm{inf}^2}{4} \frac{F'''(z_\textrm{inf})}{F'(z_\textrm{inf})} \gtrsim -1,
\end{equation}
must be checked ($n$ prime marks denote the $n$-th derivative of $F$ with respect to $z$). If this inequality holds, the inversion problem is well-posed for any amplitude and, therefore, the force deconvolution is valid. 
If not, the next step is to check whether the chosen amplitude meets the condition
\begin{equation}
	\label{eq:Awell}
	A \lesssim L_\textrm{inf} \quad \textrm{or} \quad  A \gtrsim z_\textrm{inf}/2.
\end{equation}
Here, $L_\textrm{inf} = \sqrt{-F'(z_\textrm{inf}) / F'''(z_\textrm{inf})}$ and quantifies the length scale for variation in the $F(z)$ curve. In case Eq. (\ref{eq:Awell}) is satisfied the force deconvolution is valid, otherwise, it maybe ill-posed and the amplitude must be adjusted according to Eq. (\ref{eq:Awell}) to obtain a reliable force curve. Instead of Eq. (\ref{eq:S}), also the condition for $A$ can be checked directly since Eq. (\ref{eq:S}) is derived from this condition. If the amplitude chosen in the measurement leads to a potentially ill-posed force deconvolution and the measurement cannot be repeated with a properly selected amplitude, only the force-distance curve at $z \gtrsim z_\textrm{inf} - A$ is reliably deconvoluted. 
In case the force-distance curve contains more than one inflection point, the test must be repeated for each inflection point in descending order of $z$ since the force deconvolution might only be valid to the next inflection point at a smaller $z$ value. 
\subsection{Practical Implementation}
The inflection point test requires the first and third derivative of $F(z)$ at each inflection point $z_\textrm{inf}$. However, for experimental data, both the determination of inflection points as well as the calculation of $F'(z)$ and $F'''(z)$ are not trivial due to the discrete nature of the data and noise. Small, arbitrary jumps between subsequent data points cause a set of fake inflection points (especially, for $F \approx 0$ at $z \gg 0$) and each numerical derivative of such jumps greatly amplifies the noise. To overcome these issues, the data must be filtered. Here, we decided to use MATLAB and smooth the force curve with a smoothing spline fit.\cite{MATLABss} This function fits piecewise polynomials to the data, controlled by a smoothing parameter $p$, and has three advantages compared to a running average or Gaussian filter. First, the shape of the force curve doesn't get distorted like when averaging over a few pixels so that extrema keep their magnitude. Second, since the smoothed curve consists of a set of polynomials, the function \texttt{differentiate} can calculate the first and second derivative of the force curve by simply calculating the analytical derivative of the polynomials, without creating additional noise. Third, the function \texttt{feval} allows the resolution in $z$ to be easily increased by a factor of 10 to enhance the accuracy. The third derivative of $F(z)$ is determined from another smoothing spline fit to $F'(z)$ using the same $p$ value and evaluating its second spatial derivative. The parameter $p$ must be selected by the user in a way that the fit in a small region around the tested inflection point [see Fig. \ref{fig:CO}(b)] is as smooth as possible (since this curve will be differentiated three times) but still reflects the data without any distortion. Consequently, the inflection point positions must be estimated by eye at first. With a chosen $p$ value, the code determines their exact $z$ positions automatically by looking for sign changes in $F''(z)$ for all $z$ values smaller than the position of the steepest slope in the smoothed force curve which usually excludes fake inflection points in the noise. If there is more than one inflection point and the fit to $F(z)$ is not sufficiently good for all inflection points, separate smoothing parameters $p$ must be used for each test of each inflection point. For this case (and if the described method doesn't detect all inflection points), the code allows to manually restrict the $z$ range. Finally, the inflection point test can be performed.  MATLAB code for this smoothing, differentiation and implementation of the inflection point test is available in the supplementary material. 

\section{DEMONSTRATION OF THE INFLECTION POINT TEST}
\subsection{Experimental details}

In the following, we demonstrate how to perform the inflection point test utilizing two experimental, i.\,e. discrete data sets: $\Delta f(z)$ spectra taken with a monoatomic metal tip and various amplitudes over the center of (1) a carbon monoxide (CO) molecule and (2) a single Fe trimer, both on Cu(111). Experiments were carried out with a custom-built, combined scanning tunneling and atomic force microscope operating at a temperature of 5.9\,K. A qPlus sensor\cite{Giessibl1998} equipped with an etched bulk tungsten tip, $f_0 =  20452$\,Hz (20438\,Hz for the second data set) and $k = 1800$\,N/m was used and operated in frequency-modulation mode.\cite{Albrecht1991} 
The amplitude was calibrated by recording the change in the $z$ position of the piezo when changing the oscillation amplitude in constant-current feedback mode.\cite{Simon2007,Peronio2016} A bias of $-1$\,mV ($-0.5$\,mV for the second data set) was applied to the tip. Less than 0.01 monolayers of CO molecules were dosed on a Cu(111) sample which has been cleaned by repeated sputter and anneal cycles in advance. After this, a similar amount of single Fe adatoms was evaporated onto the surface and Fe trimers were created by their lateral manipulation.

\subsection{CO/Cu(111) data set - a well-posed example for all amplitude setpoints of $A=10$\,pm, $A=30$\,pm, $A=50$\,pm, $A=75$\,pm and $A=100$\,pm}

\begin{figure}
	\includegraphics{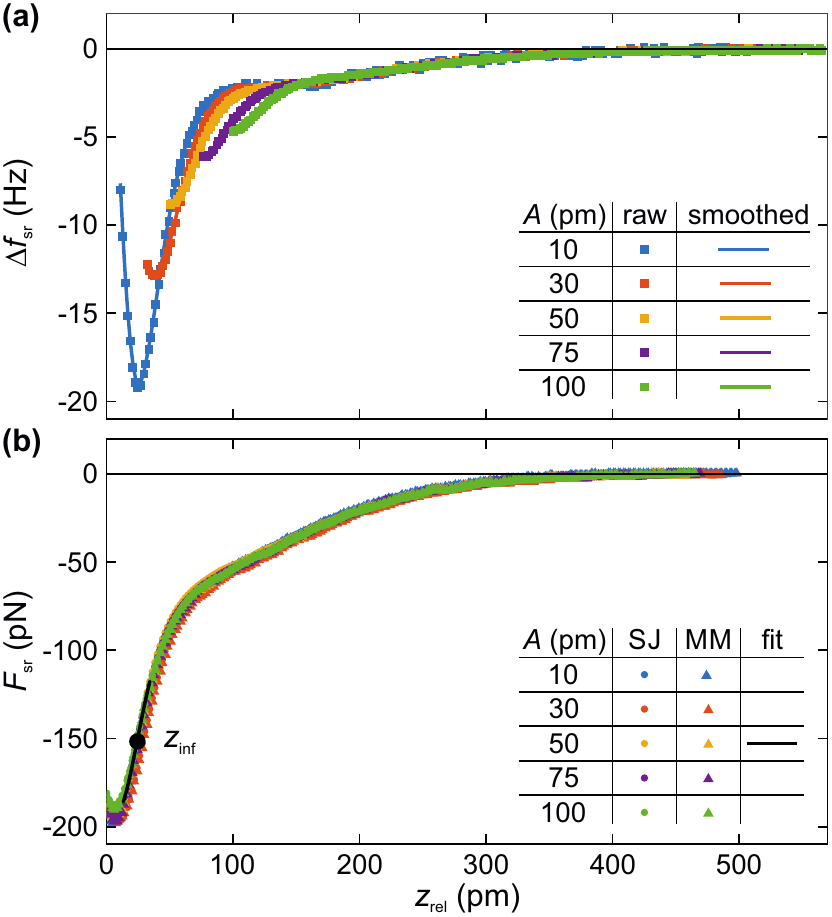}
	\caption{\label{fig:CO} Example of a well-posed force-deconvolution problem. (a) Short-range frequency shift $\Delta f$ vs. distance $z$ spectra over the center of a CO molecule for five different amplitudes. The squares indicate raw data points and the solid lines smoothed curves through these points. (b) Short-range force $F(z)$ curves calculated by a deconvolution of the corresponding data in (a) for both the Sader-Jarvis (SJ) and the matrix method (MM). For both plots, the $z$ axis is identical and set in way that the lower turnaround points of the tip oscillation of all spectra coincide in the same $z$ position (see text for all details).}
\end{figure}

The first example is a data set taken over the center of a CO molecule adsorbed on Cu(111). Figure \ref{fig:CO}(a) shows short-range $\Delta f(z)$ curves, i.e. the differences between $\Delta f(z)$ spectra over the center of the CO molecule and over the bare Cu(111) surface,\cite{Lantz2001} for five different amplitudes.
For each amplitude, the $z$ position of the tip was adjusted  prior to each measurement such that the turnaround points of the tip oscillation close to the sample of all spectra coincide in the same $z$ position defined as $z_\textrm{rel} = 0$ in Fig. \ref{fig:CO}. Any $z$ drift was compensated in the post-processing by calculating the absolute tip-sample distance based on the conductance and relating all measurements with each other. In order to suppress noise in the force-distance curves, all $\Delta f_\textrm{sr}(z)$ curves were smoothed using MATLAB's smoothing spline fit\cite{MATLABss} prior to the force deconvolution [solid lines in Fig. \ref{fig:CO}(a)]. In addition, since both the Sader-Jarvis and matrix methods require $\Delta f = 0$ at the largest $z$ value (otherwise, this will cause deconvolution errors),\cite{Welker2012b} but the corresponding $\Delta f_\textrm{sr}$ values exhibit an offset for all experimental data due to noise, these offsets were subtracted. Furthermore, for the matrix method, all $\Delta f_\textrm{sr}(z)$ curves were interpolated to a finer spacing $d$ between subsequent data points ($A/d = 50$) to avoid known numerical artifacts\cite{Welker2012b} and down-sampled again after deconvolution.

Figure \ref{fig:CO}(b) shows the short-range $F(z)$ curves as deconvoluted from the smoothed $\Delta f_\textrm{sr}(z)$ curves by the Sader-Jarvis and matrix method for five different amplitudes, respectively. The $z$ axis is identical to Fig. \ref{fig:CO}(a). Although five different amplitudes with their different particular $z$ starting points were used, all force-distance curves overlap. This is as expected by theory since the tip probed the force $F(z)$ within the same $z$ range, just with different amplitudes and, therefore, a different amplitude weighting according to Eq. (\ref{eq:df}). Small discrepancies between the curves arise from possible lateral offsets when re-positioning the tip over the molecule and deconvolution errors which can be up to 8\,\%.\cite{Sader2004,Welker2012b,Dagdeviren2018}

Selecting the force-distance curve measured with $A = 50$\,pm in Fig. \ref{fig:CO}(b), the inflection point test at the curve's single inflection point $z_\textrm{inf} = 24$\,pm  yields a $S$-factor of $-0.86$, i.\,e. a value larger than $-1$. Therefore, the force deconvolution is well-posed for any amplitude. This is also seen in the overlapping of all five $F_\textrm{sr}(z)$ curves with different amplitudes and proven by inflection point tests for all curves which all yield well-posed behavior for all chosen amplitudes.

\subsection{Fe trimer/Cu(111) data set - well posed for $A=10$\,pm and $A=100$\,pm, ill-posed for 
$A=30$\,pm and $A=50$\,pm}
\begin{figure}
	\includegraphics{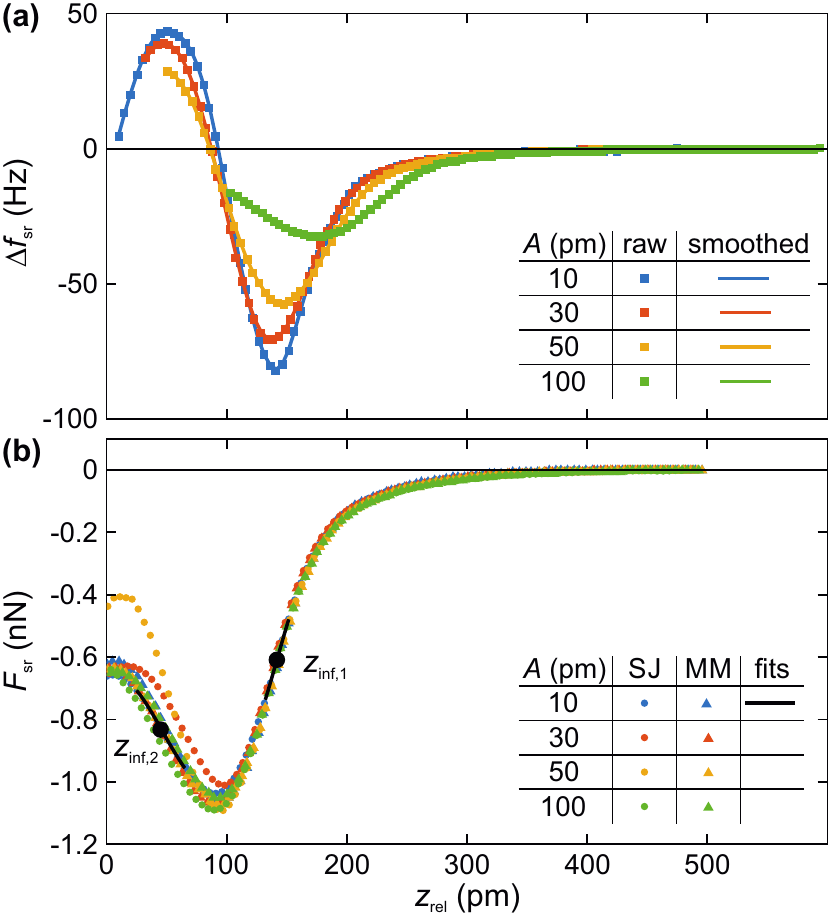}
	\caption{\label{fig:Fetri} Example of a partially ill-posed force deconvolution problem. (a) $\Delta f_\textrm{sr}(z)$ spectra over the center of a Fe trimer for four different amplitudes. (b) $F_\textrm{sr}(z)$ curves calculated by a deconvolution of the corresponding data in (a) for both the Sader-Jarvis and the matrix method. For the whole figure, data representation and $z$ axis is identical as in Fig. \ref{fig:CO}.}
\end{figure}

A different situation is given for short-range spectra over the center of a single Fe trimer on Cu(111). Figure \ref{fig:Fetri}(a) shows the short-range frequency shift vs. distance curves measured with four different amplitudes.
Data processing and $z$ axis definition are done in analogy to before. While the $F_\textrm{sr}(z)$ curves in Fig. \ref{fig:Fetri}(b) are almost identical for $z > 120$\,pm, their shapes start to deviate for smaller distances. The biggest differences are given for the $F_\textrm{sr}(z)$ curves deconvoluted with the Sader-Jarvis method for the amplitudes of 30 and 50\,pm: its curve measured with 30\,pm shows an offset compared to the other curves, its curve measured with 50\,pm an almost twice as big force gradient and an about 230\,pN weaker force at $z = 15$\,pm compared to the rest which exceeds known deconvolution errors.\cite{Sader2004,Welker2012b,Dagdeviren2018} All $F_\textrm{sr}(z)$ curves exhibit two inflection points. Consequently, the inflection point test must be started at the outermost inflection point. For the curve derived from the measurement with $A = 50$\,pm, the latter is located at $z_\textrm{inf,1} = 143$\,pm. The test for this point yields $S(F) = -7.1$ which is smaller than $-1$ pointing to potential ill-posedness per Eq. (\ref{eq:S}). According to Eq. (\ref{eq:Awell}), the force deconvolution is only well-posed if the oscillation amplitude is outside the range between $27$ and $71$\,pm. In fact, the force curves derived from the measurements within that amplitude range ($30$ and $50$\,pm) show a very different shape than the curves outside this range ($10$ and $100$\,pm). They are the result of an ill-posed force deconvolution and those curves cannot be trusted for $z < z_\textrm{inf} - A$. Therefore, a second test at the closer inflection point at $z \approx 50$\,pm is superfluous.

Following the inflection point test result, an amplitude chosen outside of the stated range should lead to well-posed behavior. The inflection point test of the $10$\,pm force-distance curve at the outermost inflection point $z_\textrm{inf,1}$ again leads to a very similar amplitude range of $26$ to $71$\,pm for possible ill-posedness. Since the amplitude is now chosen outside this range, its force-distance curve is reliable also left from this inflection point and the next inflection point at $z_\textrm{inf,2} = 46$\,pm can be tested. This results in a $S$-factor of $-0.76$ indicating that the force deconvolution is well-posed for any oscillation amplitude. Indeed, both the $F_\textrm{sr}(z)$ curves derived from the measurement with 10 and 100\,pm match each other for both the Sader-Jarvis and the matrix method for the whole $z$ range of the spectra. We note that, for this specific case, the matrix method returns virtually the same force-distance for all given amplitudes (as long as the spacing between the data points is sufficiently small).\cite{Welker2012b} Other measurements show the matrix method exhibiting stronger variations with amplitude than the Sader-Jarvis method when the inversion is ill-posed.\cite{Sader2018} Thus, it is always important to avoid force reconstruction when the inversion is ill-posed, regardless of the chosen deconvolution method. The inflection point test provides users with the means to achieve this goal.
\section{Summary}
In summary, we have explored the inflection point in detail which allows one to discriminate well- and possibly ill-posed behavior of the force deconvolution. On the basis of two examples, we have demonstrated how to apply this test to discrete experimental data. MATLAB code which semi-automates the tests and was also used in this work is available in the supplementary material. Finally, we note that a correct amplitude and piezo calibration is crucial and that the inflection point test might yield oscillation amplitudes which do not maximize the signal-to-noise ratio.\cite{Giessibl1999} However, these amplitudes should be used nonetheless in order to reliably measure forces with frequency-modulation atomic force microscopy.

See supplementary material for the MATLAB file \texttt{ifptest.m}.

We thank John Sader for helpful discussions and the Deutsche Forschungsgemeinschaft for funding within research Project No. CRC 1277, project A02.

\end{document}